\documentclass[10pt,a4paper,twocolumn]{article}
\usepackage{graphicx}
\usepackage[explicit]{titlesec}
\usepackage{authblk}
\usepackage{etoolbox}
\usepackage[labelfont=bf,labelsep=endash]{caption}
\usepackage{balance}
\usepackage{tabularx}
\usepackage{xcolor}
\usepackage[hidelinks]{hyperref}
\usepackage[hang]{footmisc}
\usepackage[normalem]{ulem}
\usepackage[top=1in, bottom=1in, left=0.69in, right=0.69in]{geometry}
\usepackage[T1]{fontenc}
\usepackage{newtxmath,newtxtext}
\usepackage{mathtools}
\usepackage[none]{hyphenat}
\usepackage{cite}
\providecommand{\keywords}[1]{\textbf{Keywords} - #1}

\titleformat{\section}{\center\normalfont\bfseries}{\thesection.}{1em}{\MakeUppercase{#1}}
\titlespacing*{\section}{0pt}{12pt}{9pt}

\titleformat{\subsection}{\normalfont\bfseries}{\thesubsection}{1em}{#1}
\titlespacing*{\subsection}{0pt}{12pt}{9pt}

\titleformat{\subsubsection}{\normalfont\itshape}{\thesubsubsection}{1em}{#1}
\titlespacing*{\subsubsection}{0pt}{12pt}{9pt}

\newcommand{\ITUurl}[1]{\textcolor{blue}{\urlstyle{same}\url{#1}}}

\def\starttable{\vspace{6pt}\begin{table}[ht]\center}
\def\startfigure{\vspace{6pt}\begin{figure}[ht]\center}

\makeatletter
\def\tagform@#1{\maketag@@@{\ignorespaces#1\unskip\@@italiccorr}}
\makeatother

\title{\large{\textbf{{Real-World Applications of AI in LTE and 5G-NR Network Infrastructure}}}}

\author{
\begin{tabular}{c}
\textbf{Simran Saxena}\hspace{3.3cm}\textbf{Arpad Kovesdy}\\
\texttt{\hspace{0.8cm}simran.saxena@beamlink.io, arpad@beamlink.io} \\
Beamlink \\
Los Angeles, USA \\
\end{tabular}
}

\date{}
\begin{document}

%%= produce title =%%
\maketitle

%%= abstract text � substitute here! =%%
\begin{abstract}
\textit{Telecommunications networks generate extensive performance and environmental telemetry, yet most LTE and 5G-NR deployments still rely on static, manually engineered configurations. This limits adaptability in rural, nomadic, and bandwidth-constrained environments where traffic distributions, propagation characteristics, and user behavior fluctuate rapidly. Artificial Intelligence (AI), more specifically Machine Learning (ML) models, provide new opportunities to transition Radio Access Networks (RANs) from rigid, rule-based systems toward adaptive, self-optimizing infrastructures that can respond autonomously to these dynamics. This paper proposes a practical architecture incorporating AI-assisted planning, reinforcement-learning-based RAN optimization, real-time telemetry analytics, and digital-twin-based validation. In parallel, the paper addresses the challenge of delivering embodied-AI healthcare services, educational tools, and large language model (LLM) applications to communities with insufficient backhaul for cloud computing. We introduce an edge-hosted execution model in which applications run directly on LTE/5G-NR base stations using containers, reducing latency and bandwidth consumption while improving resilience. Together, these contributions demonstrate how AI can enhance network performance, reduce operational overhead, and expand access to advanced digital services, aligning with broader goals of sustainable and inclusive network development.}
\end{abstract}
%%======%%

%%= keywords =%%
\begin{center}
\end{center}
%%======%%

%%= start of sectioning � modify each title and label as you please =%%
\section{Introduction} 
\label{sec:intro}
The evolution of modern telecommunications systems is increasingly shaped by the integration of artificial intelligence into network design and operation. Although LTE and 5G-NR infrastructures generate vast quantities of telemetry, from user mobility traces to channel-quality indicators and interference measurements, operational practices remain largely manual. Engineers continue to perform planning, provisioning, and optimization through labor-intensive workflows that do not effectively exploit the data already available within deployed networks. As a result, networks exhibit limited responsiveness to environmental shifts and user-behavior changes, reducing performance and increasing operational cost.

These limitations are particularly evident in rural or underserved regions, temporary event deployments, emergency-response zones, and industrial environments characterized by nomadic or seasonal usage patterns. In such settings, user density may vary unpredictably, the radio environment can change sharply over time, and network operators often lack the staff required to continually adjust configuration parameters. Static engineering assumptions rapidly become outdated, and manual interventions cannot scale to the pace or scale required.
Simultaneously, the global expansion of AI applications, including AI applications like healthcare tools, interactive educational platforms, and large language model services like OpenAI’s ChatGPT and others, poses additional challenges. These applications typically depend on cloud infrastructure for inference, which restricts accessibility in communities with limited or unreliable backhaul. Even when radio coverage is available, insufficient backhaul capacity prevents users from benefiting from advanced services. Furthermore, hosting these AI applications in datacenter incurs enormous computing and electricity costs. A potential solution: Edge hosting of AI-based applications directly on radio equipment offers an alternative pathway for delivering digital services.

ITU DataHub statistics highlight the magnitude of this challenge: Mobile-phone ownership exceeds 75–90\% of the population in most regions and continues to rise, even in lower-income economies, increasing the number of active devices competing for limited radio resources \cite{ITU2023MobileOwnership}. Meanwhile, global mobile broadband subscriptions have surpassed 110 subscriptions per 100 inhabitants, confirming that multi-SIM usage and device proliferation further intensify RAN load \cite{ITU2023Subscriptions}, \cite{ITU2024Subscriptions}. Monthly mobile data consumption continues to grow at double-digit annual rates, contributing to rising congestion in areas where infrastructure growth lags behind demand \cite{itu_datahub}.
These limitations are particularly evident in rural or underserved regions, temporary-event deployments, emergency-response zones, environments affected by nomadic or seasonal usage changes, and industrial environments. In many countries, ITU Datahub reports show rural 4G population coverage below 60\%, despite mobile-phone ownership exceeding 85\%, demonstrating a clear imbalance between user capacity and available infrastructure \cite{itu_datahub}, \cite{ITU2025MobileCoverage}. Static engineering assumptions rapidly become outdated, and manual interventions cannot scale to the pace or scale required.

This paper discusses these two problem domains by presenting an architecture for AI-driven RAN optimization alongside an edge-hosting model for AI applications. The resulting system supports both autonomous network operation and accessible digital services in resource-constrained regions. Furthermore, Beamlink, a company founded in 2017 in the United States to democratize construction of cellular network infrastructure, implements many of the discussed research in real-world products operating globally. In this paper, the generalized Beamlink architecture, which consists of Bentocells (all-in-one integrated base stations with compute capability) that can exist as femtocells (0.25W), microcells (1-5W), and macrocells (50W+) interface with cloud resources (like Maia, Beamlink's cloud controller), however, the same interfaces can be implemented in other paradigms like OpenRAN, or with other proprietary ecosystems from other vendors.

\section{Existing Infrastructure Challenges}
\label{sec:sec2}
The first challenge motivating this work is the intrinsic rigidity of manual RAN planning. Engineers must determine transmit power, antenna tilt and azimuth, scheduling strategies, handover thresholds, QoS policies, and interference-mitigation parameters, often using planning tools like Planet, IBWave, or Forsk Atoll that are not directly linked to the RAN controllers or cellular infrastructure. These decisions are often based on predicted user distributions and assumed propagation conditions that fail to capture real-world variability. With mobile-phone ownership increasing globally and mobile data traffic growing at more than 20–30\% CAGR in many regions \cite{EricssonDataTrafficGrowth}, networks configured using static assumptions show degradation in quality of service, uneven load distribution, and operational inefficiencies.

Monitoring and tuning live networks pose a second challenge. RAN telemetry includes RSRP and SINR measurements, interference estimates, bit error rates, HARQ statistics, congestion indicators, dropped call information, and throughput measurements. In LTE and 5G New Radio (NR), Reference Signal Received Power (RSRP) and Reference Signal Received Quality (RSRQ) are standardized downlink metrics representing the received signal strength and quality of reference signals, respectively \cite{3gpp_ts_36_214} \cite{3gpp38_300}. RSRP is defined as the linear average power of resource elements carrying cell-specific reference signals, while RSRQ indicates the signal quality relative to the total received wideband power (including interference and noise) \cite{3gpp_ts_36_214}. Additionally, link-level indicators such as Hybrid Automatic Repeat Request (HARQ) feedback and the Channel Quality Indicator (CQI) reflect transmission reliability and instantaneous channel conditions for modulation adaptation, respectively \cite{3gpp38_300}. These telemetry parameters are commonly used as state features in reinforcement learning-based network optimization algorithms to enhance coverage and performance \cite{drone}.
Although these metrics reflect the real-time health of the network, interpreting them requires expertise and significant manual effort. Network operators often rely on periodic audits rather than continuous evaluation, making optimization reactive rather than proactive. ITU data shows that cell traffic distribution has become increasingly bursty as mobile-broadband adoption expands, further straining manual processes \cite{ITU2025InternetTraffic}.

A third challenge concerns the delivery of AI-driven digital services. Telemedicine, assistive robotics, rehabilitation systems, and interactive educational platforms increasingly rely on cloud-hosted AI models, including high-bandwidth vision systems or large-scale language models. These services cannot function reliably in regions with limited backhaul capacity. ITU reports show that affordability and bandwidth constraints remain major barriers in developing economies, where the cost of a 2 GB mobile-broadband plan still exceeds recommended affordability thresholds \cite{ITU2023Affordability}, \cite{ITUPriceBrief2023}. Even when local radio coverage exists, constraints in upstream connectivity effectively prevent access to advanced digital tools.
Together, these challenges motivate the need for RAN intelligence capable of autonomously adapting to dynamic conditions, as well as mechanisms for running AI applications locally at the network edge.

\section{AI-enabled Network Planning}
\label{sec:sec3}
AI-driven planning provides a means of replacing static engineering assumptions with learned models that can infer optimal configurations based on historical datasets and real-world measurements. In rural deployment scenarios, for example, where 100 radios may be required to fill a coverage gap, user density can shift dramatically throughout the day or season. ML (Machine Learning) models can estimate optimal transmit power, antenna orientation, beam patterns, and channel assignments by analyzing mobility traces, terrain characteristics, and interference patterns. Unlike rule-based approaches, these models improve over time as more data becomes available.

Machine learning architectures such as Graph Neural Networks (GNNs) can represent the RAN as a structured graph, in which nodes correspond to base stations and edges represent interference or backhaul relationships. By learning these interactions, GNNs can predict coverage boundaries, identify regions of excessive overlap, and propose cell-level corrections. Clustering models similarly help uncover latent spatial patterns in user behavior, allowing the network to adjust proactively as demand shifts. 

In parallel, reinforcement learning approaches such as Q-learning provide a complementary paradigm in which the network learns optimal control policies directly through interaction with the environment, rather than relying on explicitly engineered input–output mappings. By observing state transitions and long-term rewards (e.g., throughput, interference reduction, or handover stability), Q-learning agents can adapt transmission power, handover thresholds, or scheduling decisions in large-scale LTE and 5G-NR deployments. Clustering and unsupervised models further uncover latent spatial and temporal patterns in user behavior that are difficult or impossible to identify manually, enabling proactive network reconfiguration as traffic demand evolves.

Reinforcement learning (RL) introduces the possibility of continuous adaptation. RL agents treat RAN configuration as a sequential decision-making problem in which actions such as adjusting antenna tilt or modifying power levels are evaluated through a reward function reflecting network performance objectives. Early studies demonstrate potential energy savings through automatic MIMO-layer management, as well as improved latency and throughput through adaptive scheduling strategies \cite{erol-kantarci}.

Recent work has explored complementary dimensions of RAN energy-efficiency and performance trade-offs by explicitly manipulating physical layer and networking parameters to reduce power consumption. For example, studies have analyzed the impact of dynamic MIMO configuration—including selectively enabling or disabling MIMO modes based on traffic or channel conditions—to reduce base station power draw, using value functions tied to eNodeB utilization as input features for self-optimization frameworks that balance throughput and energy use \cite{minimize_base_station_power}.
Other research has quantified timing constraints in packet transfers as operational constraints in joint communication-control optimization, where metrics such as maximum allowed delay (MAD) and maximum allowable transfer intervals (MATI) are incorporated into optimization problems to ensure ultra-low latency or reliability while minimizing communication power costs \cite{ali}
Additionally, foundational optimization frameworks for energy-aware wireless networks demonstrate that turning off underutilized antennas or access points and planning network flexibility with energy management in mind can substantially reduce overall RAN energy consumption without violating quality of service (QoS) requirements, informing design choices where capacity, delay, and power efficiency are jointly considered \cite{energy}.

Beyond high-level configuration control, recent work has examined reinforcement-learning formulations that directly optimize low-level physical-layer resource allocation \cite{subband_control}. In this research, learning-based approaches have been proposed for joint sub-band selection and transmit-power control, addressing the mixed discrete–continuous nature of spectrum allocation while operating with delayed or locally available channel state information rather than instantaneous global Channel State Information (CSI). These methods replace centralized, model-driven optimization—such as fractional programming—with distributed deep reinforcement learning architectures, combining deep Q-learning for discrete sub-band scheduling and actor–critic methods for continuous power control. Such RL models create even further improvements especially in dense and interference-limited networks where spectrum is valuable.

By using reinforcement learning models, base station properties, beamforming patterns, antenna systems, entire access points, and even sub-band control can be adjusted using the feedback gained from reward and penalty systems in neural networks to save countless engineering hours adjusting and debugging base station parameters. Beamlink utilizes these RL modeling systems using a combination of the Bentocell radio plus processing architecture at the edge, as well as Beamlink's cloud network controller called Maia to collect information from base stations and the network core to make real-time and periodic adjustments.

\section{Real-Time RAN Monitoring and Optimization}
\label{sec:sec4}
Once deployed, RAN sites generate fine-grained telemetry reflecting their operational state. This includes the spatial distribution of RSRP and SINR values, time-varying interference levels, HARQ retransmission statistics, scheduler behavior, and throughput performance across user equipment. AI-based systems can process this telemetry in real time to detect anomalies, identify coverage deterioration, or predict emerging congestion. For example, unsupervised learning techniques can identify deviations from established patterns of network behavior, enabling early detection of hardware failures or misconfigurations.

Predictive models extend these capabilities by forecasting traffic spikes or mobility surges, allowing the network to allocate resources proactively rather than reactively. When combined with reinforcement learning, such models support closed-loop optimization, in which configuration parameters are continually tuned based on real-time feedback.

A digital twin provides a controlled environment for safely evaluating these adaptations. By maintaining a virtual replica of the live RAN that evolves based on continuous telemetry input, operators can test new parameters or train RL agents without risking service disruptions. The digital twin simulates interference interactions, traffic dynamics, and propagation characteristics, allowing for pre-deployment validation of AI-generated configuration updates. By interacting with a simulated or digital-twin environment, an RL agent can learn control policies without risking live network degradation.

\startfigure
\includegraphics[width=\columnwidth]{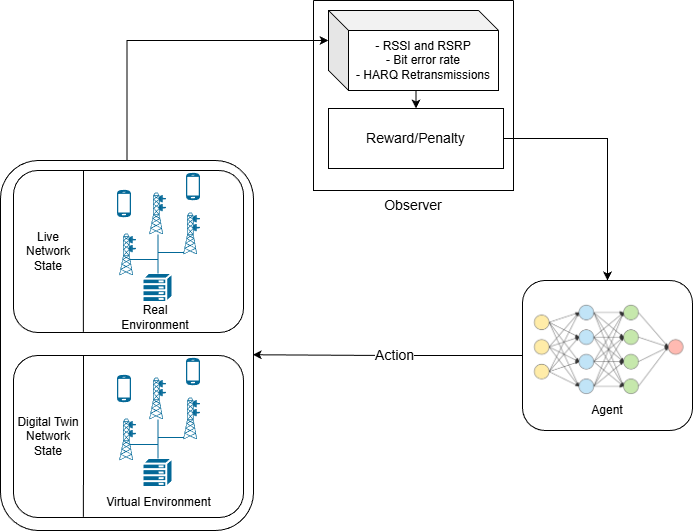}
\caption{ML agents take actions to modify the settings of network infrastructure, either in a real environment or a virtual environment. An observer software program measures key statistics and produces rewards and penalties for the agent in a cycle.}\label{fig:fig1} 
\end{figure}

In practical deployments, the effectiveness of real-time RAN monitoring and optimization depends on access to fine-grained telemetry and programmable control interfaces. Beamlink's flagship base station Bentocell provides an example of a software-defined base station architecture that exposes detailed PHY- and MAC-layer measurements, including RSRP and SINR distributions, HARQ statistics, per-UE scheduling metrics, interference estimates, noise power, and throughput measurements. These data are continuously collected across operational deployments and can be used to construct datasets suitable for anomaly detection, traffic prediction, and reinforcement-learning-based control.
In addition, Maia aggregates telemetry across distributed base stations and provides mechanisms for both monitoring and control. Maia interfaces with radio and core-network configuration components, enabling automated adjustment of radio parameters, scheduling policies, and mobility-related settings. This bidirectional data and control path supports the implementation of closed-loop optimization workflows, in which AI-generated configuration updates can be evaluated and applied. Together, the availability of detailed telemetry and remote configuration capabilities enables the platform to serve as a testbed for studying real-time RAN optimization, digital-twin-assisted validation, and the deployment of learning-based control algorithms in operational cellular environments.

\section{System Architecture for Closed-loop RAN Control}
\label{sec:sec5}
\startfigure
\includegraphics[width=\columnwidth]{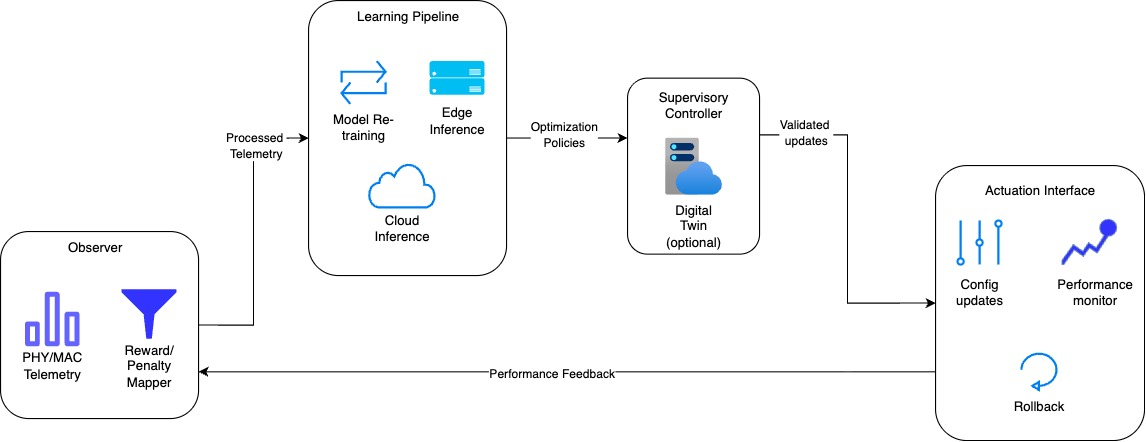}
\caption{High-level overview of the System Architecture for Closed-Loop RAN control.}\label{fig:fig2} 
\end{figure}
The proposed system implements a closed-loop control architecture for LTE and 5G-NR Radio Access Networks in which sensing, learning, validation, and actuation are continuously integrated. A data-collection layer, referred to as the \textit{observer}, aggregates fine-grained PHY and MAC telemetry from base stations, including RSRP, SINR, HARQ statistics, buffer occupancy, scheduler states, and throughput metrics. These measurements are standardized within 3GPP specifications and are therefore readily available for AI-based processing \cite{3gpp38_300}. To reduce dimensionality and stabilize learning, the observer can map raw telemetry into abstract reward and penalty signals reflecting operator objectives such as spectral efficiency, latency, fairness, and energy consumption.

Learning pipelines periodically retrain models using historical and real-time data, enabling adaptation to evolving traffic patterns, mobility dynamics, and propagation conditions. Inference may be executed either at the network edge or in centralized cloud infrastructure depending on latency constraints and computational availability. Fast control loops, such as scheduling and power adjustment, favor edge-based inference, while slower optimization tasks such as coverage tuning and parameter re-provisioning are well suited to centralized execution.

An actuation interface applies validated configuration updates to live RAN nodes, including power levels, antenna parameters, and scheduler weights. To prevent destabilizing actions, all updates pass through a supervisory controller that enforces safety constraints derived from regulatory limits, equipment capabilities, and operator policies. Performance is continuously monitored after actuation, and rollback mechanisms restore prior configurations if degradation is detected, consistent with ETSI ENI recommendations for supervised autonomy \cite{etsi_eni_005}.

To further reduce operational risk, the architecture supports integration with a digital twin of the RAN. Candidate policies can be evaluated against a virtual replica driven by real telemetry before deployment, enabling reinforcement-learning agents to explore optimization strategies without impacting live service. The resulting closed-loop system enables continuous, autonomous adaptation to changes in user behavior, traffic demand, and environmental conditions, reducing reliance on manual optimization while improving scalability and resilience.

\section{Edge-Hosted Applications and Embedded AI Services}
\label{sec:sec6}
Beyond improving radio performance, AI also influences how digital services can be delivered in resource-constrained environments. Cloud-based AI workloads, such as real-time video analytics, telemedicine diagnostics, or LLM (Large Language Model) inference, often require high-throughput backhaul links that are unavailable in many rural regions. Hosting these applications directly on LTE/5G-NR base stations allows users to interact with advanced services without relying on remote servers.
This model supports a wide range of applications. For example, in an example private LTE deployment where cameras are streaming security footage through an LTE network, back to a cloud application that analyzes the footage, camera streams from IoT devices can be analyzed locally on the same hardware running eNodeB or gNodeB functions, with only relevant alerts transmitted externally. This could reduce the cloud costs of a company operating security systems and AI monitoring of those security cameras, for example. In addition, the backhaul necessary may be reduced from above 100 Mbps to a negligible value if only network traffic is generated when an event occurs (like a possible intruder enters the area).

Another example is Embedded-AI healthcare applications. Imagine AI applications that could help diagnose common illnesses, analyze radiography (X-rays, CT-scans), or retrieve past patient history securely. However, when Internet backhaul is constrained, for example, if a hospital is undergoing a natural disaster and backhaul has been severed or overloaded with users, these critical applications will also be disconnected just when the hospital, and its doctors and nurses, need them the most. In these crucial emergencies, hosting the applications locally on the eNodeBs and gNodeBs or nearby telecommunications hardware can provide an unprecedented level of reliability. In addition, the latency during normal operation would be substantially better than if the applications were hosted remotely. Hospitals, clinics, and other healthcare facilities would see no difference between remotely hosted and edge-hosted applications, other than better reliability during backhaul outages and better latency of requests.

\startfigure
\includegraphics[width=\columnwidth]{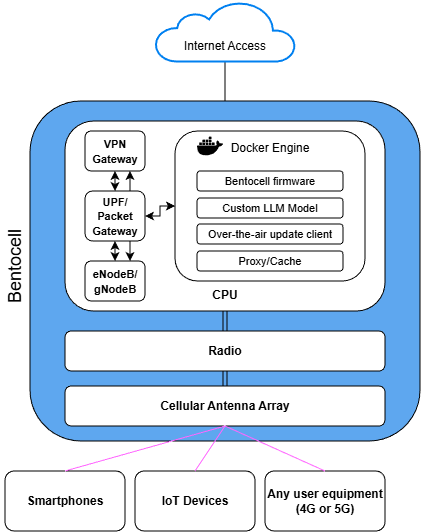}
\caption{Hosted application architecture on a Bentocell (4G/5G radio with additional hosted application support) using Docker Engine to host applications in secure containers, networked to an onboard packet gateway.}\label{fig:fig3} 
\end{figure}

Finally, a third example is that educational content and interactive tools can be cached at the edge and served to hundreds of students simultaneously. In many communities, a school may have hundreds to thousands of students who need access to educational content, like learning videos, assessments, textbooks, etc. If each time a student watches a video or accesses a resource requires the use of a backhaul connection, this is not a problem for many schools located in cities, but in rural areas with weak or non-existent connectivity, this is a major bottleneck. Cellular infrastructure that can be used for the dual purpose of establishing connectivity to end-user devices like phones, tablets, and computers, but also caches content like educational videos, textbooks, or even small LLMs that can be interacted with by students, can enable efficient learning (low-latency) while conserving bandwidth for occasional usage or material updates at off-hours. Even systems like language translation for language learning can be hosted locally, allowing for new learning avenues to be opened for all students, not just those in cities.

Overall, implementing edge hosting requires support for containerized execution environments (like hosting a custom Docker container on edge hardware as shown in Figure 2), isolated virtual network functions, and local storage for models and cached content. Modern software-defined network infrastructure like Beamlink’s virtualized RAN hardware already incorporates sufficient processing capability to support such workloads, making this architecture feasible for near-term deployments. The Bentocell platform integrates all functional elements required for both radio access and edge computation within a single, compact base-station architecture. In addition to RF and baseband processing, each Bentocell includes a general-purpose processing subsystem that traditionally implements centralized and distributed unit (CU/DU) functions using heterogeneous compute resources such as CPUs, GPUs, and optional NPUs, coupled with local RAM and persistent storage. High-speed interfaces, including Ethernet and fiber backhaul, connect the system to core networks while also enabling local data-plane services shown in Figure 3. This convergence allows the same hardware platform to host external applications alongside RAN functions, including embedded AI workloads, reinforcement-learning agents for network optimization, and auxiliary services such as content caches or proxy servers. By co-locating compute, storage, and radio functions at the network edge, the Bentocell enables low-latency decision-making, reduced backhaul dependency, and tighter integration between AI-driven control loops and live network operation. Using over-the-air updates, hypervisors and container hosting systems can be deployed, even to eNodeB’s and gNodeB’s operating in the field, and new hardware may be designed with newer processors that contain NPUs (Neural Processing Units) to make machine learning inference more efficient in hosted applications. As a result, the architecture supports both adaptive RAN optimization and the localized delivery of digital services within a unified deployment model, to allow for better service in specific situations like for healthcare, learning, and industry, but also general-purpose data service.

\section{Conclusion}
\label{sec:sec10}
This paper presents a unified approach to integrating AI-driven RAN optimization with edge-hosted AI services. By replacing static engineering practices with adaptive, data-driven models, networks become more resilient, efficient, and responsive. Edge hosting further ensures that communities without reliable backhaul connectivity can access modern digital services, including healthcare, education, and AI-driven tools. Together, these innovations support the development of sustainable, inclusive communication infrastructures that align with future visions of intelligent network operation.

\bibliographystyle{IEEEtran}
\bibliography{references}

\end{document}